\documentclass{caosp}
\usepackage{graphicx}
\articleNo{230}
\pubyear{2014}
\volume{44}
\volnumber{2}
\firstpage{1}
\received{March 28, 2014}
\accepted{2014}
\begin{document}
\title{Supernova 2014J at maximum light}
\author{
        D.Yu.\,Tsvetkov \inst{1}
   \and
       V.G.\,Metlov \inst{1}   
   \and
        S.Yu.\,Shugarov \inst{1,2}
   \and
        T.N.\,Tarasova \inst{3}
   \and
        N.N.\,Pavlyuk \inst{1}
}
   \institute{
          Sternberg Astronomical Institute, M.V.\,Lomonosov Moscow State
          University,
          Universitetskii pr. 13, 119992 Moscow, Russia\\
           \email{tsvetkov@sai.msu.su}
         \and
           \lomnica
         \and
          Crimean Astrophysical Observatory, Nauchnyi, Crimea}

\hauthor{D.Yu.\,Tsvetkov {\it et al.}}
\date{March 19, 2014}
\maketitle
\begin{abstract}
We present {\it UBVRI} photometry of the supernova 2014J in M82,
obtained in the period from January 24 until March 3, 2014, as well as two
spectra, taken on February 4 and March 5.
We derive dates and magnitudes of maximum light in the {\it UBVRI} bands,
the light curve parameters $\Delta m_{15}$ and expansion velocities of the
prominent absorption lines.
We discuss colour evolution, extinction and maximum luminosity of SN 2014J.

\end{abstract}

\section{Introduction}

Supernova (SN) 2014J, located at $\alpha=9^{\rm h}55^{\rm m}42^{\rm s}.14,
\delta=+69^{\circ}40'26''.0$ (2000.0) in the galaxy M82, 
was discovered by Steve J. Fossey
on UT 2014 January 21.8. The description of discovery and early 
observations were presented by Goobar {\it et al.} (2014).
The prediscovery observations and early spectra were also 
reported by Zheng {\it et al.} (2014).
These sets of data
show that SN 2014J is a spectroscopically normal Type Ia SN,
although it exhibited high-velocity features in the spectrum and
was heavily reddened by the dust in the host galaxy.

At a distance of 3.5 Mpc (Karachentsev and Kashibadze, 2006)
SN 2014J is the nearest SNIa since SN 1972E, and it offers the unique
possibility to study a thermonuclear SN over a wide range of the 
electromagnetic spectrum. 

\section{Observations}

We present here CCD photometry of SN 2014J in the
{\it UBVRI} passbands obtained at three sites. Nearly daily coverage
was achieved in the period from January 24 until March 3, 2014.
Observations were carried out at the Crimean Observatory of the Sternberg
Astronomical Institute (SAI)(Nauchnyi, Crimea); at the Moscow Observatory of 
SAI (Moscow, Russia) and at the Star\'a Lesn\'a Observatory
of the Astronomical Institute of the Slovak Academy of Sciences.

A list of the observing facilities is given
in Table\,\ref{t1}.

\begin{table}
\begin{center}
\caption{Telescopes and detectors employed for the observations.}\vskip2mm
\label{t1}
\begin{tabular}{ccccccc}
\hline\hline
Tele-  & Location &
Aperture &  CCD  & Filters & Scale & FoV  \\
scope &  & [m] & camera  &  &   [arcsec & [arcmin] \\
 code  &  &     &  &  &     pixel$^{-1}$]             &     \\
\hline
S60  & Star\'a        & 0.6 & FLI     & $UBVR_CI_C$ &0.85 & 14.0 \\
     & Lesn\'a,       &     & ML 341  &          &      &    \\
     & Slovakia       &     &         &          &      &   \\
K50  & Nauchnyi,      & 0.5 & Apogee  & $UBVR_CI_C$& 1.10 & 30.5x23.0 \\
     & Crimea         &      &Alta U8300  &         &      &     \\
M70  & Moscow,        & 0.7 & Apogee  & $UBVR_CI_J$& 0.64 & 5.5 \\
     & Russia         &     &  AP-7p  &          &      &     \\
M20  & Moscow,        & 0.2 & Apogee  & $UBVR_CI_J$  & 1.22 & 10.4 \\
     & Russia         &     & AP-7p   &          &      &      \\
\hline\hline
\end{tabular}
\end{center}
\end{table}

The standard image reductions and photometry were made using the
IRAF\footnotemark .
\footnotetext{IRAF is distributed by the National Optical 
Astronomy Observatory,
which is operated by AURA under cooperative agreement with the
National Science Foundation.}
The magnitudes of the SN
were derived by a PSF-fitting relatively to
two bright local standard stars.
The surface brightness of the host galaxy at the location of the SN
is quite high, and subtraction of galaxy background is necessary 
for accurate photometry. We had no images of M82 obtained at our instruments
before SN outburst, and used the images downloaded from the CASU
archive\footnotemark .\footnotetext{http://casu.ast.cam.ac.uk}
They were transformed to match our images using appropriate IRAF tasks.

\begin{figure}
\centerline{\includegraphics[width=11cm,clip=]{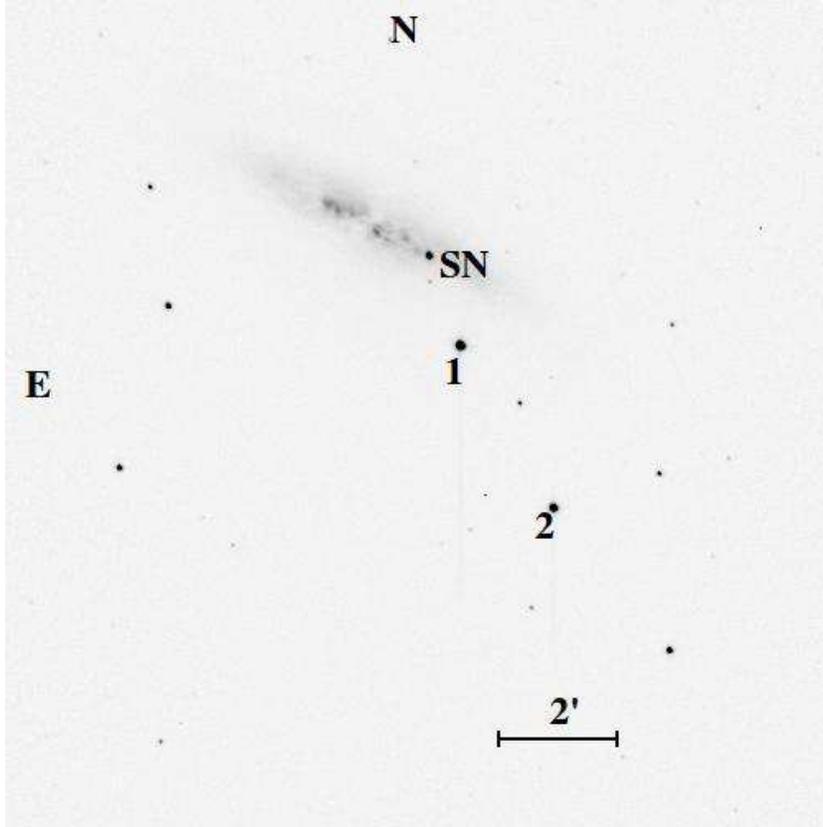}}
\caption{The image of SN 2014J and local standard stars, obtained
at the S60 telescope in the $V$-band on February 23.}
\label{f1}  
\end{figure}

The CCD image of SN 2014J and local standard stars is presented
in Fig.\,\ref{f1}.

The magnitudes of the local standards were calibrated on 7 nights
relative to a standard in the field of the nearby galaxy M81 
(Richmond {\it et al.}, 1996),
they are reported in Table\,\ref{t2}. 
The $B,V$-band magnitudes of star 1 are in a good agreement with the
data reported by AAVSO\footnotemark ,
\footnotetext{http://www.aavso.org/download-apass-data}
but for the star 2 the difference between our results and AAVSO data
is significant. 

\begin{table}
\begin{center}
\tabcolsep=5pt
\caption{$UBVRI$ magnitudes of local standard stars}
\label{t2}
\begin{tabular}{ccccccccccc}
\hline\hline
Star  & $U$ & $\sigma_U$ & $B$ & $\sigma_B$ & $V$ & $\sigma_V$ &
$R$ & $\sigma_R$ & $I$ & $\sigma_I$\\
\hline
 1 & 10.86 & 0.02& 10.62 & 0.01 & 10.04 & 0.01 & 9.70 &  0.01&  9.41 &  0.01 \\
 2 & 12.17 & 0.02& 11.53 & 0.01 & 10.70 & 0.01 & 10.23&  0.01&  9.85 &  0.02 \\
\hline\hline
\end{tabular}
\end{center}
\end{table}

The photometry was transformed to the standard Johnson-Cousins
system by means of instrument colour-terms, determined from
observations of standard star clusters. The procedure was described
in details by Elmhamdi {\it et al.} (2011) and Tsvetkov
{\it et al.} (2008).
The type of $R$ and $I$-filters is indicated in Table\,\ref{t1}.
We transformed the photometry in the $R, I$-bands to Cousins
system, so $R$ and $I$ are equivalent to $R_C$, $I_C$.

The photometry of the SN is presented
in Table\,\ref{t3}.

\begin{table}
\begin{center}
\tabcolsep=5pt
\caption{$UBVRI$ magnitudes of SN2014J}
\label{t3}
\begin{tabular}{cccccccccccc}
\hline\hline
JD$-$ & $U$ & $\sigma_U$ & $B$ & $\sigma_B$ & $V$ & $\sigma_V$ &
$R$ & $\sigma_R$ & $I$ & $\sigma_I$& Tel. \\
2456000 & & & & & & & & & & &\\
\hline
682.13&       &     & 12.49 & 0.03 & 11.20 & 0.03 & 10.52&  0.03&  10.03&  0.03&  M20\\
686.20& 13.00 & 0.12& 12.08 & 0.02 & 10.77 & 0.03 & 10.24&  0.03&   9.77&  0.03&  M20\\
687.13& 12.90 & 0.16& 11.99 & 0.03 & 10.70 & 0.02 & 10.16&  0.02&   9.72&  0.02&  M20\\
688.17& 13.02 & 0.25& 11.98 & 0.03 & 10.67 & 0.02 & 10.14&  0.02&   9.72&  0.02&  M20\\
689.13& 12.93 & 0.18& 11.95 & 0.03 & 10.61 & 0.02 & 10.11&  0.03&   9.71&  0.03&  M20\\
691.18& 12.72 & 0.04& 11.85 & 0.03 & 10.56 & 0.01 & 10.05&  0.01&   9.86&  0.02&  K50\\
692.20& 12.79 & 0.04& 11.88 & 0.03 & 10.57 & 0.02 & 10.06&  0.02&   9.90&  0.02&  K50\\
693.17& 12.79 & 0.03& 11.91 & 0.03 & 10.58 & 0.01 & 10.08&  0.01&   9.93&  0.02&  K50\\
694.17& 12.73 & 0.05& 11.94 & 0.04 & 10.59 & 0.02 & 10.08&  0.02&   9.95&  0.02&  K50\\
694.49& 12.88 & 0.05& 12.01 & 0.04 & 10.65 & 0.02 & 10.11&  0.02&   9.99&  0.03&  S60\\
695.18&       &     & 11.99 & 0.03 & 10.61 & 0.02 & 10.12&  0.02&   9.99&  0.02&  K50\\
696.16&       &     & 12.03 & 0.03 & 10.63 & 0.02 & 10.17&  0.02&  10.02&  0.02&  K50\\
700.17& 13.06 & 0.08& 12.29 & 0.02 & 10.82 & 0.02 & 10.42&  0.02&  10.27&  0.02&  K50\\
701.20&       &     & 12.39 & 0.02 & 10.89 & 0.02 & 10.52&  0.02&  10.32&  0.02&  K50\\
702.30& 13.46 & 0.04& 12.48 & 0.03 & 10.95 & 0.02 & 10.60&  0.01&  10.37&  0.03&  K50\\
702.45& 13.53 & 0.04& 12.55 & 0.03 & 10.94 & 0.01 & 10.57&  0.02&  10.38&  0.02&  S60\\
707.17& 14.09 & 0.05& 13.00 & 0.03 & 11.25 & 0.02 & 10.81&  0.01&  10.39&  0.02&  K50\\
708.22&       &     & 13.19 & 0.05 & 11.28 & 0.03 & 10.77&  0.04&       &      &  M20\\
708.23& 14.29 & 0.07& 13.08 & 0.02 & 11.29 & 0.02 & 10.81&  0.02&  10.37&  0.02&  K50\\
711.35& 14.60 & 0.06& 13.40 & 0.04 & 11.41 & 0.02 & 10.81&  0.02&  10.29&  0.03&  K50\\
712.18&       &     & 13.49 & 0.06 & 11.45 & 0.01 & 10.81&  0.02&  10.27&  0.03&  K50\\
712.25& 14.85 & 0.06& 13.56 & 0.04 & 11.43 & 0.02 & 10.76&  0.02&  10.21&  0.02&  S60\\
714.22&       &     & 13.80 & 0.04 & 11.51 & 0.02 & 10.84&  0.02&  10.13&  0.03&  M20\\
714.40& 14.99 & 0.05& 13.71 & 0.03 & 11.47 & 0.02 & 10.80&  0.02&  10.12&  0.02&  S60\\
715.18&       &     & 13.84 & 0.04 & 11.54 & 0.02 & 10.85&  0.02&  10.11&  0.02&  M20\\
715.42& 15.12 & 0.05& 13.79 & 0.02 & 11.53 & 0.02 & 10.81&  0.01&  10.13&  0.02&  S60\\
716.31& 15.49 & 0.04& 14.04 & 0.02 & 11.63 & 0.01 & 10.90&  0.01&  10.14&  0.01&  M70\\
716.41& 15.26 & 0.07& 13.90 & 0.04 & 11.58 & 0.03 & 10.85&  0.03&  10.12&  0.02&  S60\\
717.23& 15.48 & 0.04& 14.13 & 0.03 & 11.67 & 0.01 & 10.94&  0.02&  10.16&  0.02&  M70\\
718.42& 15.34 & 0.04& 14.08 & 0.03 & 11.65 & 0.01 & 10.86&  0.02&  10.08&  0.02&  S60\\
720.18& 15.42 & 0.05& 14.11 & 0.03 & 11.81 & 0.02 & 10.94&  0.01&  10.18&  0.02&  K50\\
\hline\hline         
\end{tabular}        
\end{center}         
\end{table}

The spectroscopic observations were carried out at the 2.6-m Shajn telescope
of CrAO on 2014 February 4 and March 5. Spectrograph SPEM provided 
the wavelength range of 3300--7550 \AA\,  with a
dispersion of 2\AA\, pixel$^{-1}$. The spectra were bias and
flat-field corrected, extracted and wavelength calibrated with
the SPERED code developed by S.I.Sergeev at the Crimean Astrophysical 
Observatory.
The spectrophotometric standard HR3894 was used for flux calibrated
spectra.

\section{Light and colour curves}

\begin{figure}
\centerline{\includegraphics[width=11cm,clip=]{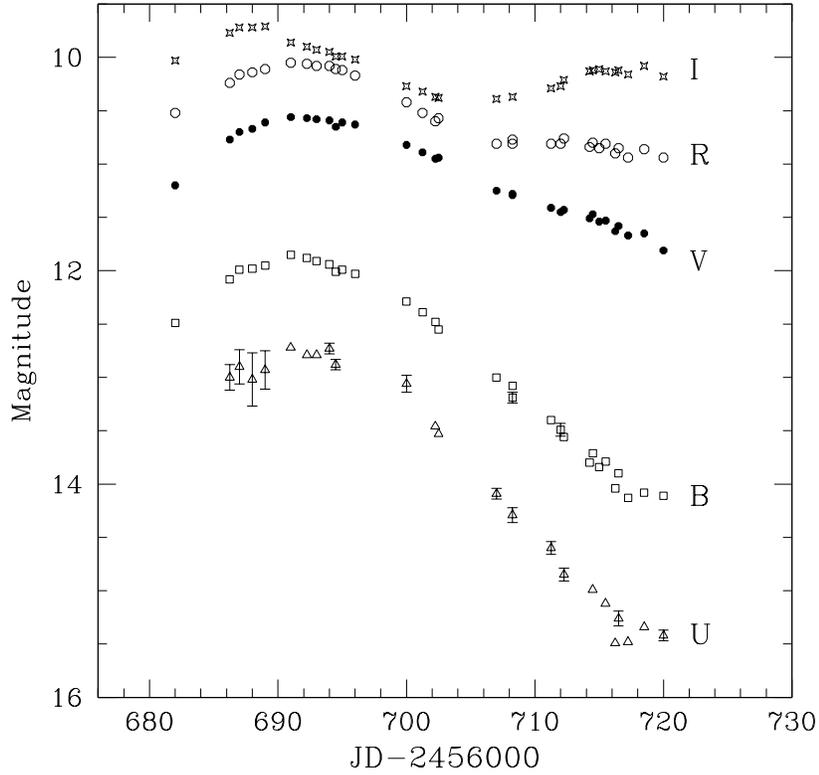}}
\caption{The light curves of SN 2014J in the {\sl UBVRI} bands.
The error bars are plotted only when they exceed the size of
a symbol.}
\label{f2}
\end{figure}

The light curves of SN 2014J are presented in Fig.\,\ref{f2}.
The results for all the telescopes are in a
fairly good agreement, some systematic differences can be noted 
only for the magnitudes in the $U$ and $I$-bands.
The shape of the light
curves is typical for SNe Ia.

We fitted the light curves with cubic splines and determined the
dates and magnitudes of maximum light in different bands and the
decline rate parameters $\Delta m_{15}$. These data
are reported in Table\,\ref{t4}.
The values of $\Delta m_{15}$ confirm that SN 2014J is a normal type Ia
SN with rate of brightness decline close to the mean values. 
The explosion likely occurred on January 14.72 (Zheng {\it et al.}, 2014, 
Goobar {\it et al.}, 2014)(JD 2456672.22), and we can determine that the time
interval between explosion and the $B$-band maximum equals 19.2 days.  

\begin{table}
\begin{center}
\caption{Dates and magnitudes of maximum light and the
decline rate parameters in different passbands.}
\label{t4}
\begin{tabular}{cccc}
\hline\hline
Band & JD$-$2456000   & mag & $\Delta m_{15}$ \\
\hline
$U$  & 691.5$\pm$ 2.0 & 12.76$\pm$ 0.14 &  1.27$\pm$ 0.15 \\
$B$  & 691.4$\pm$ 0.4 & 11.88$\pm$ 0.05 &  1.01$\pm$ 0.05 \\
$V$  & 691.9$\pm$ 0.3 & 10.57$\pm$ 0.04 &  0.63$\pm$ 0.04 \\
$R$  & 691.5$\pm$ 0.5 & 10.04$\pm$ 0.04 &  0.72$\pm$ 0.05 \\
$I$  & 687.7$\pm$ 1.5 &  9.71$\pm$ 0.12 &  0.68$\pm$ 0.10 \\
\hline\hline
\end{tabular}
\end{center}
\end{table}

\begin{figure}
\centerline{\includegraphics[width=11cm,clip=]{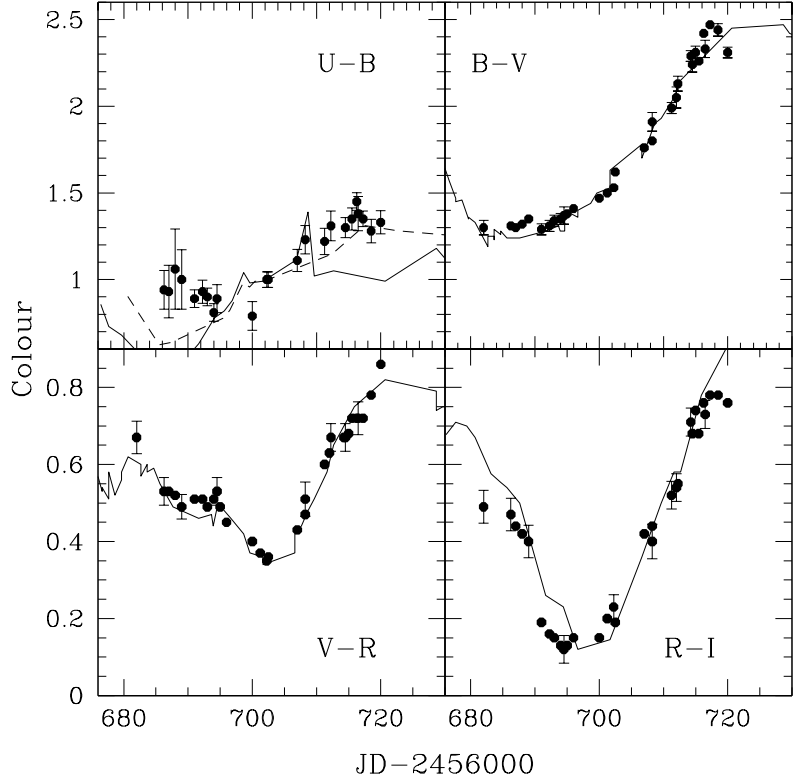}}
\caption{The colour curves of SN 2014J. The solid lines present 
colour curves for SN 2011fe, and dashed line is for the $(U-B)$
curve for SN 2003du.}
\label{f3}
\end{figure}

The colour curves for SN 2014J are presented in Fig.\,\ref{f3}.
The colour evolution is typical for SN Ia, this is confirmed by
comparison with the colour curves for SN 2011fe (Tsvetkov {\it et al.}, 2013),
which is a "normal", unreddened SN Ia
with nearly the same $\Delta m_{15}$.  
The evolution of all colours, except $(U-B)$, is similar for the two
objects. The behaviour of the $(U-B)$ colour is different for SNe 
2014J and 2011fe. We plotted also the $(U-B)$ color curve for SN Ia
2003du (Stanishev {\it et al.}, 2007), it is a better match for the curve of
SN 2014J, but the differences are still evident. 
The amount of shift applied to match the curves can be considered as
an estimate of 
the colour excess of SN 2014J. 
We obtained the following estimates: $E(U-B)=1.1\pm 0.15$; 
$E(B-V)=1.3\pm0.05$; $E(V-R)=0.47\pm0.05$; $E(R-I)=0.6\pm0.05$.
The colour excess due to the galactic extinction is $E(B-V)_{gal}=0.14$
according to Schlafly and Finkbeiner (2011), so the colour excess
in the host galaxy is $E(B-V)_{host}=1.16$. Assuming distance modulus for
M82 $\mu=27.73$ (distance 3.5 Mpc, Karachentsev and Kashibadze 2006) 
and mean absolute magnitude
for SN Ia with $\Delta m_{15}=1.01$ $M_B=-19.35$ (Prieto {\it et al.}, 2006)
we obtain for ratio of total to selective extinction $A_V/E(B-V)\approx1.5$.
This number is much smaller than the typical galactic value of 3.1,
but close to the values found for another heavily reddened type Ia 
SNe (see e.g. Wang {\it et al.}, 2008).   

\section{Spectra}

\begin{figure}
\centerline{\includegraphics[width=11cm,clip=]{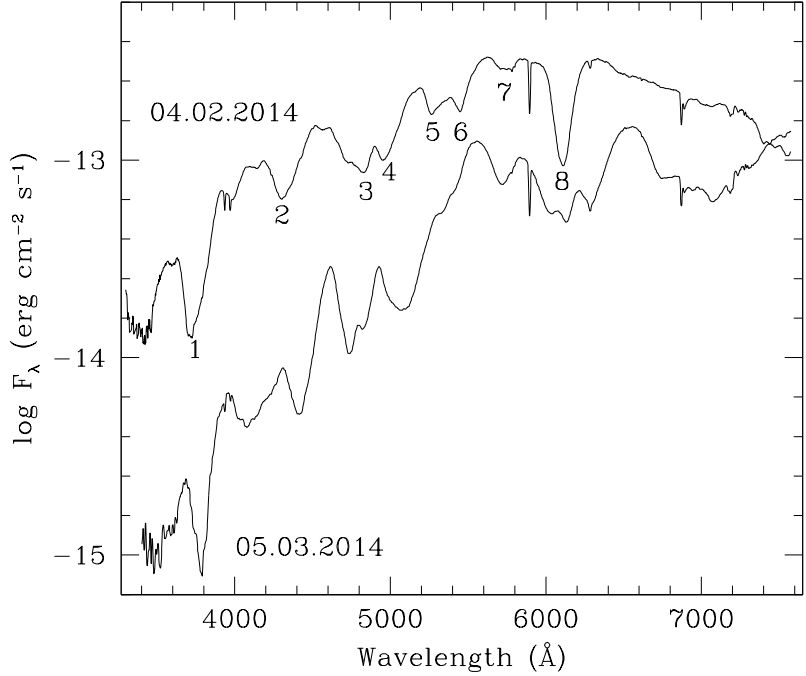}}
\caption{Spectra of SN 2014J. The identification of strongest absorption 
features: 1: CaII H\&K, SiII $\lambda$3858; 2: MgII $\lambda$4481,
FeII $\lambda$4404; 3,4: blends of several lines of FeII, FeIII, SiII;
5: SII $\lambda$5454; 6: SII $\lambda$5640; 7: SiII $\lambda$5972;
8: SiII $\lambda$6355.}
\label{f4}
\end{figure}

The spectra of SN 2014J obtained at the 2.6-meter telescope on February 4
(phase 2 days after the $B$-band maximum)
and March 5 (phase 30 days) are shown
in Fig.\,\ref{f4}.
The spectra are typical for SNe Ia at corresponding phases.
 
We estimated the expansion velocities from the wavelengths of 
prominent un-blended absorption features and corrected them
for the radial velocity of M82. For the phase 2 days
we obtain $v$=11620 km s$^{-1}$ for the line SiII$\lambda$6355,
$v$=10330 km s$^{-1}$ for the line SII$\lambda$5640.
For the epoch 30 days we find $v$=10780 km s$^{-1}$ for the 
line SiII$\lambda$6355.
The velocities are in good agreement with the results of Srivastav
{\it et al.} (2014), they are higher than average for SNe of this type.
The interstellar Na D line is very strong, we derive its equivelent
width $EW$(Na D)$=5.8$\AA, in agreement with the data reported
by Cox {\it et al.} (2014) and Kotak (2014).  

\section{Conclusions}

We present the light and colour curves of SN 2014J starting 9 days
before the $B$-band maximum and continuing until day 29 past
maximum. The spectra were obtained at phases 2 days and 30 days
after the $B$-band maximum.

The light and colour curves for SN 2014J show that it belongs to the
"normal" subset of type Ia SNe, but is heavily reddened by the dust
in the host galaxy.
We estimate the decline rate parameter $\Delta m_{15}(B)=1.01$ which
is close
to the mean value for SNe Ia. 
The comparison of colour excess and the luminosity, expected from
Pskovskiy-Phillips relation, results in low value of the ration of
selective to total extinction, similar to the values found for 
other highly reddened type Ia SNe. 

The spectral evolution is typical for this class of SNe, with expansion
velocities higher than the mean values.

We continue the observations of SN 2014J, the results and more detailed
analysis of the data will be presented in a subsequent paper.

\acknowledgements
The work of DT and NP was partly supported by the RFBR
grant 13-02-92119. 

This paper makes use of data obtained from the Isaac Newton Group Archive 
which is maintained as part of the 
CASU Astronomical Data Centre at the Institute of Astronomy, Cambridge.

\end{document}